\documentclass[a4paper]{jpconf}

\usepackage{graphicx}
\usepackage{booktabs}
\newcommand*{\wn}{cm$^{-1}$}

\begin{document}
\title{Constraining extra space dimensions using precision molecular spectroscopy}

\author{Beatriz Gato-Rivera$^{1,2, }$\footnote[3]{Also known as B. Gato. Talk given at the "Seventh International Workshop DICE2014
Spacetime - Matter - Quantum Mechanics", Castiglioncello, Italy, 15 - 19 September 2014.}}

\address{$^1$ Instituto de F\'\i sica Fundamental, IFF-CSIC, Serrano 123, 28006 Madrid, Spain}
\address{$^2$ Nikhef, Science Park 105, 1098 XG Amsterdam, The Netherlands}

\ead{t38@nikhef.nl}

\begin{abstract}
Highly accurate measurements of quantum level energies in molecular systems provide a test ground for new physics, as such
effects could manifest themselves as minute shifts in the quantum level structures of atoms and molecules. 
For the lightest molecular systems, neutral molecular hydrogen (H$_2$, HD and D$_2$) and the molecular hydrogen ions
(H$_2^+$, HD$^+$ and D$_2^+$), weak force effects are several orders weaker than current experimental and theoretical results,
while contributions of Newtonian gravity and the strong force at the characteristic molecular distance scale of 1 \AA\ can be
safely neglected. Comparisons between experiment and QED calculations for these molecular systems can be interpreted in terms
of probing large extra space dimensions, under which gravity could become much stronger than in ordinary 3-D space. Under this
assumption, using the spectra of H$_2$ we have derived constraints on the compactification scales for extra dimensions within
the Arkani-Hamed-Dimopoulos-Dvali (ADD) framework, and constraints on the brane separation and bulk curvature within the
Randall-Sundrum (RS-I and RS-II) frameworks.

\end{abstract}

\section{Introduction}

This presentation is based on work done in collaboration with Bert Schellekens, from the Institute of Subatomic Physics
Nikhef (Amsterdam), together with Edcel Salumbides and Wim Ubachs, from the Vrije Universiteit of Amsterdam. 
In the present study, accurate results from precision molecular spectroscopy are exploited to put constraints on some
higher-dimensional theories. The idea is that effects of new physics could manifest themselves as minute shifts in the 
quantum level structures of atoms and molecules. 
To detect such phenomena, spectroscopic experiments must be performed at extreme precision, where key ingredients are
ultra-stable lasers and laser metrology techniques, and comparisons must be made on systems for which very accurate
first-principles calculations can be performed.

In recent years great progress has been made on the calculation of level energies in the neutral hydrogenic molecules 
(H$_2$, HD and D$_2$) and their ions (H$_2^+$, HD$^+$ and D$_2^+$). 
The results from the variety of experimental precision measurements on both the ionic and neutral hydrogenic molecules
are in perfect agreement with the QED-calculations, within combined uncertainty limits from theory and experiment. 
This agreement between experiment and first-principles calculations on the level energies of hydrogenic molecules can be 
used to constrain the parameters associated with Physics Beyond the Standard Model, such as possible fifth forces between
nucleons or the existence of extra space dimensions.
Since weak, strong and (Newtonian) gravitational forces have negligible contributions to the molecular quantum level
structure, electromagnetism is the sole force acting between the charged particles within light molecules. This makes it
possible to derive bounds on possible fifth forces and extra dimensions from molecular precision experiments compared with
QED-calculations~\cite{Salumbides2013,Salumbides2014}.

Although in the framework of String Theory extra space dimensions were found to be necessary from the very begining
in the middle seventies of last century, it took much longer to appreciate the beauty and 
usefulness of extra dimensions in ordinary Quantum Field Theory. Only in the late nineties, after the seminal works of
Antoniadis, Arkani-Hamed, Dimopoulos, and Dvali (ADD model)~\cite{Hamed1998,Hamed1998b,Antoniadis} on one side, and Randall and Sundrum 
(RS-I and RS-II models)~\cite{Randall1999,Randall1999b} on the other, the particle physics community started considering extra 
dimensions as a real scientific possibility deserving further investigation. The mathematical formalisms of these models
can be applied to molecular physics test bodies, from which constraints on the compactification scales can be derived 
for the ADD scenario, whereas constraints on the brane separation and bulk curvature can be derived for the RS-I and RS-II 
scenarios.

\section{The ADD Scenario}

In the model put forward in 1998 by Arkani-Hamed, Dimopoulos, and Dvali~\cite{Hamed1998,Hamed1998b} 
(plus Antoniadis~\cite{Antoniadis} in its implementation in string theory) our observable universe consists 
of a 3-brane embedded in a higher-dimensional cosmos. There are $n$ extra space dimensions where only gravity 
can escape as the particles and interactions of the Standard Model remain confined in the ordinary 3-D space 
(up to a small, negligible penetration inside the bulk). The extra dimensions are flat and compactified with the
same radius $R$, just for simplicity. Therefore the extradimensional space is an $n$-dimensional torus where gravitation 
propagates rendering the observed gravitational strength much weaker as the flux lines dilute into the higher-dimensional 
volume. Moreover, the propagation of gravity across the extradimensional space results in a modification of Newton's
law. The exact ADD gravitational potential reads~\cite{Hamed1998b}:

\begin{equation}
  V_\mathrm{ADD}(r) =  - G  \frac{m_1m_2}{r} \sum_{(k_1,k_2,...k_n)}  e^{(2\pi |k|/R)r}      
\label{ExactADD}
\end{equation}

It has the contribution from the usual massless gravitons plus a tower of Yukawa potentials mediated by all the massive
modes: the Kaluza-Klein higher-dimensional gravitons, with momenta in the extra dimensions quantized in units of
$2\pi /R$, and therefore with mass splittings of $2\pi /R$ from our 3-D space perspective (from the higher-dimensional
perspective they are massless). For distances much larger than the compactification length range, $r \gg R$, the Newtonian 
potential is recovered:

\begin{equation}
  V_\mathrm{ADD}(r)= V_\mathrm{N}(r) = - G  \frac{m_1m_2}{r} = - \frac{\hbar c}{M_\mathrm{Pl}^2} \frac{m_1m_2}{r} ,  \quad  r \gg R
\end{equation}                                               
where $M_\mathrm{Pl}$ is the Planck mass ($10^{16}$ TeV). For distances of the same size or shorter than the 
compactification length range, $r \leq R$, the ADD potential takes the simple form:

\begin{equation}
 V_\mathrm{ADD}(r)= - G_{(4+n)}\frac{m_1m_2}{r^{n+1}} = - \frac{\hbar c}{M_{(4+n)}^{(n+2)}} \frac{m_1m_2}{r^{n+1}} ,  \quad  r \leq R
\label{VADD}
\end{equation}
where $M_{(4+n)}$ is the fundamental mass in the (4+$n$)-dimensional spacetime. 
Observe that this expression for the gravitational potential can be interpreted 
in the context of the Gauss law, without the need to invoke quantization of gravity. The Planck mass $M_\mathrm{Pl}$ is then 
related to the fundamental higher-dimensional mass $M_{(4+n)}$ via:
\begin{equation}
  M_\mathrm{Pl}^2 =  M^{n+2}_{(4+n)} (R)^n.
\label{Fund_M_Pl}
\end{equation}
Thus the fundamental mass $M_{(4+n)}$ may still be small and $M_\mathrm{Pl}$ becomes large due to the compactified
volume of extra dimensions. Arkani-Hamed {\it et al.} proposed that $M_{(4+n)}$ equals the electroweak scale
$M_{EW}$ in order to solve the hierarchy problem, explaining in this manner the desert between the electroweak scale 
of 1 TeV and the Planck scale of $10^{16}$ TeV. In the present study, since it is not our intention to solve the hierarchy problem,  
we leave $M_{(4+n)}$ as a free parameter, together with $R$ and $n$, and we attempt to constrain these 
parameters from molecular physics experiments.

The gravitational potential has an effect on the level energy of a molecular quantum state with wave function $\Psi(r)$, 
written as an expectation value~\cite{SSGU}:
\begin{equation}
\left<V_\mathrm{ADD}\right> = G  m_1m_2 \left[ \quad \int_{R}^{\infty} \Psi^*(r) \frac{1}{r} \Psi(r) r^2 dr  + 
R^n \int_0^{R} \Psi^*(r) \frac{1}{r^{n+1}} \Psi(r) r^2 dr \quad \right]
\label{Integrals}
\end{equation}

\begin{figure}
\begin{center}
\resizebox{0.48\textwidth}{!}{\includegraphics{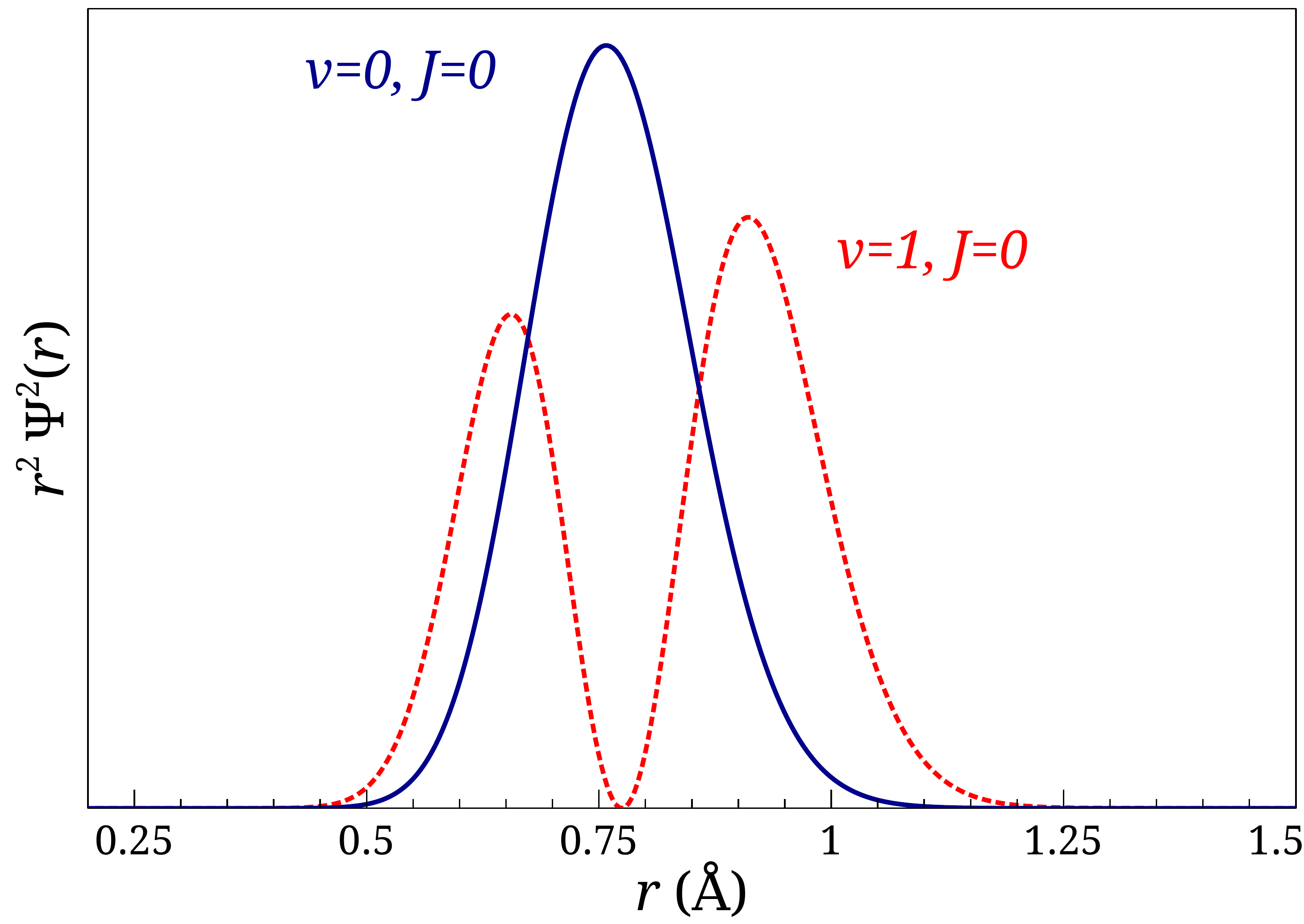}}
\end{center}
\caption{(Color online) Wave functions for H$_2$ in the electronic ground state with $J=0$ and vibrational levels $v=0, 1$.}
\label{Wavefunctions}
\end{figure}

$\Psi(r)$ represents the probability that the internuclear separation is $r$, the vibrational coordinate that probes the
forces between the nucleons. The first integral represents the ordinary gravitational attraction, 
which for protons is $8 \times 10^{-37}$ times weaker than the electrostatic repulsion, and can therefore be neglected.
The second integral represents the effect of modified gravity and is evaluated using accurate wave functions for the 
molecules. The wave functions of the H$_2$ ground electronic state for $J=0$ and vibrational levels $v=0, 1$ are plotted in
Fig.~\ref{Wavefunctions} as obtained from \emph{ab initio} calculations~\cite{Piszcziatowski2009,Komasa2011}. 
In practice, the integration is performed up to $r=10$ \AA\ since the wave function amplitude is
negligible beyond that. In addition, at distances $r < 0.1$ \AA\ the wave function amplitude becomes negligible so that the 
second integral in Eq.~(\ref{Integrals}) converges without additional assumptions.

For spectroscopic transitions between quantum states $\Psi_1$ and $\Psi_2$ in molecules, the expectation value for a
higher-dimensional gravity contribution is given by a differential effect:
\begin{equation}
 \left<\Delta V_\mathrm{ADD}(n,R) \right> =  G  m_1m_2 R^n \times 
\left[ \left< {\Psi_1} \left| \frac{1}{r^{n+1}} \right|{\Psi_1} \right> - \left< {\Psi_2} \left| \frac{1}{r^{n+1}} \right|{\Psi_2} \right> \right]
\label{ADD-trans}
\end{equation}
Here the ADD-expectation value is written explicitly as a function of the two relevant parameters:
the number $n$ of extra space dimensions and the compactification scale $R$. Obviously, measurements on the 
dissociation limit of molecules, where $\Psi_1$ is the lowest energy bound state and $\Psi_2$ is the non-interacting 
two-atom limit at $r=\infty$, are the most sensitive probes as the diference between $\Psi_1$ and $\Psi_2$ is larger.

\section{The Randall-Sundrum Scenarios}

Let us now consider the Randall-Sundrum scenarios, RS-I~\cite{Randall1999} and RS-II~\cite{Randall1999b}, presented by
the authors in 1999. They consist of
two parallel 3-branes separated by a distance $y_c$ along one extra dimension $y$. The branes have opposite tension $T$
(vacuum energy) and the bulk has a negative cosmological constant. According to
General Relativity, where all types of energy are gravity sources, the branes and the bulk produce gravitational fields,
giving rise to the AdS metric:
\begin{equation}
ds^2 = e^{-2 k |y|} \eta_{\mu\nu}d x^\mu d x^\nu + dy^2,
\label{RS_metric}
\end{equation}
where $\exp(-k|y|)$ is the warp factor and $k$ is the bulk curvature. The strength of gravity is highest on the 
positive $T>0$ tension brane and falls off along the extra dimension.

In analogy with the ADD model, in the RS models the particles and interactions of the Standard Model are confined
in one 3-brane, that we denote the SM-brane, whereas gravity propagates everywhere. Observe that the warped metric
crucially differentiates the RS models from the ADD model with a flat metric where $k=0$. 

The modified gravitational potential between two masses separated a distance $r$ in the SM-brane can be expressed as:
\begin{equation}
	V_\mathrm{RS}(r) = - G \frac{m_1m_2}{r} \left( 1 + \Delta_\mathrm{RS} \right), 
\end{equation}
where $\Delta_\mathrm{RS}$ is the RS-I or RS-II correction to the Newtonian potential.

In the RS-I scenario the SM-brane is the negative $T<0$ tension brane. The authors did not present any gravitational
potential for this model, however, and it was
Callin~\cite{Callin2004b}, in 2004, who presented its computation.
For short distances, $kr \ll 1$, he obtained:
\begin{equation}
  \Delta_\mathrm{RS-I}(r) \simeq
     \frac{4}{3\pi kr}\frac{1-e^{-2k y_c}}{1+\frac{1}{3}e^{-2k y_c}},  \quad kr \ll 1,
\label{RSI_potential_short}
\end{equation}
where one can distinguish two regimes, $k y_c \ll 1$ and $k y_c \gg 1$, with the result at leading orders:
\begin{equation}
\renewcommand{\arraystretch}{2}
  \Delta_\mathrm{RS-I}(r) \simeq \left\{
  \begin{array}{l l}
   \frac{2y_c}{\pi r} + ..., & \quad ky_c \ll 1,\\
   \frac{4}{3 \pi kr} + ..., & \quad ky_c \gg 1.
  \end{array} \right.
\renewcommand{\arraystretch}{1}
\label{RSI_potential_ky}
\end{equation}
The potential for long distances,  $kr \gg 1$,  does not apply for the scales probed by molecules (it gives inconsistent results) 
and we do not consider it further.

In the RS-II scenario the SM-brane is the positive $T>0$ tension brane. The  $T<0$ tension brane can be set
to infinity, $y_c\rightarrow\infty$, as the massless gravitons are localized very close to the $T>0$ brane, their wave 
function decreasing drastically along the extra dimension $y$. Therefore, in the RS-II scenario the extra dimension
needs not be compactified. The authors derived a potential  with the long distance limit  that has been applied 
to astrophysics  (see for example~\cite{Iorio}). For short distances, $kr \ll 1$, Callin and Ravndal~\cite{Callin2004a} obtained:
\begin{equation}
  \Delta_\mathrm{RS-II}(r) \simeq
     \frac{4}{3\pi kr} + ......,  \quad kr \ll 1. 
\label{RSII_potential_short}
\end{equation}
Note the correspondence of this result with that of Eq.~(\ref{RSI_potential_ky}) in the limit  $ky_c \gg 1$.
This limit together with $kr \ll 1$ seems to indicate a smooth transition from RS-I to RS-II for small bulk curvature
$k \ll 1/r$ and $y_c \gg 1/k$, as the brane separation $y_c$ is increased to infinity~\cite{Callin2004a}.
  
From the RS corrections to the Newtonian potential the expectation values of the 
leading-order shifts of transitions in molecules,  for short distances $kr \ll 1$,  are therefore:
\begin{equation}
 \left<\Delta V_\mathrm{RS}(k) \right> =  G  m_1m_2 \mathcal{F} \left(\frac{4}{3\pi k}\right) \times 
 \left[ \left< {\Psi_1} \left| \frac{1}{r^2} \right|{\Psi_1} \right> - \left< {\Psi_2} \left| \frac{1}{r^2} \right|{\Psi_2} \right> \right],
\label{RS-trans-short}
\end{equation}
where $\mathcal{F}=(1-e^{-2k y_c})/(1+\frac{1}{3}e^{-2k y_c})$ for the RS-I scenario and $\mathcal{F}=1$ for the RS-II
scenario. Using these expressions, we derive bounds on the brane separation $y_c$ and 
the bulk curvature $k$ from molecular spectroscopy data.  

The RS-II scenario with a single brane can be generalized to $n$ extra dimensions provided that the mixing effects of the warp
factors for each extra dimension are small compared with the leading order terms. 
In this case the resulting correction to the Newtonian potential is, to a good approximation, the product of the leading corrections 
of each extra dimension. For $n$ extra dimensions, if the bulk curvatures have the same value $k$, the RS-II correction is then 
$\Delta_{\mathrm{RS-II},n} = (\Delta_\mathrm{RS-II})^n$, where we use $\Delta_\mathrm{RS-II}$ in Eq.~(\ref{RSII_potential_short}).

\section{Constraints on extra dimensions from molecular spectroscopy data}

In the previous sections, the expectation value for a higher-dimensional gravity contribution to a transition frequency in a molecule
was presented for both ADD and RS approaches to higher dimensions. This expectation value is interpreted as a contribution to the 
binding energy of molecules in certain quantum states. However, as discussed in Ref.~\cite{SSGU}, analysis of recent experiments with
hydrogen neutral molecules and their ions, together with their stable isotopomers containing deuterons, showed that QED computations
are in very good agreement with observations for these molecular systems. Consequently, one can derive constraints on the parameters
underlying the higher-dimensional theories: the compactification scale $R$ depending on the number of extra dimensions $n$, for the 
ADD scenario, and the bulk curvature $k$ or brane separation $y_c$ for the RS scenarios. For that purpose one can use the combined uncertainty  
$\delta E$ for molecular systems, where $\delta E_\mathrm{theory}$ and $\delta E_\mathrm{exp}$ are uncertainties of theory and experiment:
\begin{equation}
 \delta E = \sqrt{\delta E_\mathrm{exp}^2 + \delta E_\mathrm{theory}^2}.
\label{Uncertainties}
\end{equation}

In the ADD scenario the constraining relation
\begin{equation}
\left<\Delta V_\mathrm{ADD}(n,R)\right> \, < \delta E
\label{ADD-constrain-relation}
\end{equation} 
translates on constraints on the compactification scale $R$, depending on $n$, via:

\begin{equation}
 (R)^n < \frac{\delta E}{G  m_1m_2 \Delta} ,
\label{Rn-constraint}
\end{equation}
where $\Delta$ is the difference in expectation values over the wave function densities in the molecule:
\begin{equation}
 \Delta = \left[ \left< {\Psi_1} \left| \frac{1}{r^{n+1}} \right|{\Psi_1} \right> - \left< {\Psi_2} \left| \frac{1}{r^{n+1}} \right|{\Psi_2} \right> \right]
\end{equation}

As  shown in Fig.~\ref{Wavefunctions}, the wave functions for the lowest vibrational states, in the case of H$_2$ and for $J=0$, 
 are located in the same region of space. Consequently, the fundamental 
vibrational transition in the hydrogen molecule ($v=0 \rightarrow v=1$) probes only a differential effect.
Constraints on $R$, obtained from a comparison with the fundamental vibrational transition of H$_2$ are 
presented in Table~\ref{Constraints-R}. 

\begin{table}
\caption{Constraints on the compactification scale $R$ of extra dimensions (in units of m) as derived from both the measurement
of the fundamental ($v=0\rightarrow1$) vibration in H$_2$ and of the dissociation limit $D_0$ of H$_2$. The constraints 
are derived within the ADD-scenario assuming that the $n$ extra dimensions are of equal size. The higher-dimensional Planck length 
$R_{\mathrm{Pl},(4+n)}$ (in units of m) and Planck mass $M_{(4+n)}$ (in units of GeV) for $D_0$ are also tabulated.}
\vskip .3truecm
\label{Constraints-R}
\begin{center}
\begin{tabular}{l@{\hspace{15pt}}l@{\hspace{15pt}}l@{\hspace{15pt}}l@{\hspace{15pt}}l}
\toprule
 $n$ & $R (v=0\rightarrow1)$ & $R (D_0)$ & $R_{\mathrm{Pl},(4+n)} (D_0)$  & $M_{(4+n)} (D_0)$ \\
\hline

2 & $2.2\times10^{ 4}$ & $1.0\times10^{ 4}$ & $4.0\times10^{-16}$  & $4.9\times10^{-1}$\\
3 & $7.7\times10^{-1}$ & $1.9\times10^{-1}$ & $4.5\times10^{-15}$  & $4.4\times10^{-2}$\\
4 & $1.1\times10^{-3}$ & $8.5\times10^{-4}$ & $2.3\times10^{-14}$  & $8.7\times10^{-3}$\\
5 & $3.3\times10^{-5}$ & $3.2\times10^{-5}$ & $7.1\times10^{-14}$  & $2.8\times10^{-3}$\\
6 & $3.4\times10^{-6}$ & $3.7\times10^{-6}$ & $1.7\times10^{-13}$  & $1.2\times10^{-3}$\\
7 & $6.9\times10^{-7}$ & $7.8\times10^{-7}$ & $3.3\times10^{-13}$  & $6.0\times10^{-4}$\\

\hline
\end{tabular}
\end{center}
\end{table}

The experimental as well as the theoretical results for the fundamental vibration in the hydrogen molecule are known to the 10$^{-4}$ \wn\ 
level, an order of magnitude more accurate than the values for the binding energies~\cite{Dickenson2013,Niu2014}. However for a comparison 
of dissociation limits, it is no longer a small difference along the internuclear coordinate axis that is probed, but the difference between 
the molecular scale of 1 \AA\  and infinite atomic separation. The expectation value for the ADD-contribution to the binding energy of the
lowest bound state in the H$_2$ molecule,  the $D_0$ binding energy, is:
\begin{equation}
\left<\Delta V_\mathrm{ADD}(n,R)\right> =  G  m_1m_2 R^n \left< \frac{1}{r^{n+1}} \right>_{\Psi_0}
\end{equation}
By comparing to the experimental findings on $D_0$(H$_2$) \cite{Liu2009} this leads to another set of constraints on $R$ for $n$ extra dimensions,
which are also listed in Table~\ref{Constraints-R}. 
In Ref.~\cite{SSGU} more constraints on $R$ are presented corresponding to the dissociation limits of the deuterium molecule
$D_2$ and the ($v=0\rightarrow4$) R(2) ro-vibrational transition in HD$^+$.

Similarly, one can derive constraints pertaining to corrections in the RS models and the combined uncertainty $\delta E$ for a specific 
molecular transition
\begin{equation}
 \left<\Delta V_\mathrm{RS}(k) \right> < \delta E.
\end{equation}
Using $\delta E$ for the dissociation limit $D_0$(H$_2$) we obtain the following constraints. 
For the RS-I scenario in the short distance ($kr \ll 1$) regime, using Eq.~(\ref{RSI_potential_ky}) one finds a constraint on the brane 
separation of $y_c<2\times 10^{18}$ m in the limit $ky_c\ll 1$, and a constraint for the inverse of the curvature of $1/k < 3\times 10^{18}$ m 
in the limit $ky_c\gg 1$. 

For the RS-II scenario with one extra dimension in the short distance ($kr \ll 1$) regime, using 
Eq.~(\ref{RS-trans-short}) one finds a constraint for the inverse of the curvature of $1/k < 3\times 10^{18}$ m. 
For the case of $n>1$ extra dimensions with small curvatures with the same value $k$, using results from the $D_0$(H$_2$) 
analysis one obtains constraints on the inverse of the curvature $1/k$ depending on $n$. For $n=2$,
$1/k < 1.5\times10^{+4}$ m; for $n=3$, $1/k < 2.6\times10^{-1}$ m; for $n=4$, $1/k < 1.0\times10^{-3}$ m; for 
$n=5$, $1/k < 3.9\times10^{-5}$ m; for $n=6$, $1/k < 4.3\times10^{-6}$ m and for $n=7$, $1/k < 8.8\times10^{-7}$ m.

\section{Comparison with other constraints}

The constraints obtained from molecular systems probe length scales in the order of Angstroms.
This complements bounds probing subatomic to astronomical length scales obtained from other studies using distinct methodologies.
The micrometer to millimeter range are probed in torsion-balance type experiments, with the tightest constraint obtained by 
Kapner \emph{et al.}~\cite{Kapner2007} for a single extra dimension of $R < 4.4 \times 10^{-5}$ m.

Precision spectroscopies of hydrogen and muonic atoms have been interpreted in terms of the ADD-model~\cite{Li2007} resulting in typical
constraints of $R < 10^{-5}$ m. The interpretation is not straightforward because of the proton size puzzle~\cite{Pohl2010, Antognini2013}; 
in fact, the argument has been turned around, and the existence 
of extra dimensions are instead invoked as a possible solution to the puzzle~\cite{WangNi}. 
In addition, unlike for molecules, in the treatment of atoms some assumptions had to be made on the wave function density at $r=0$, 
typical for the $s$-states involved, causing problems in calculating the second integral of Eq.~(\ref{Integrals}) over the 
\emph{electronic} wave function that has a significant amplitude at $r=0$ in atoms.

To probe length scales in the subatomic range, one is ultimately limited by the increasing contributions from nuclear structure and the 
strong interaction. In contrast to QED calculations, the most accurate lattice-QCD calculation of light hadron 
masses only achieves relative accuracies in the order of a few percent. 

In high-energy particle collisions, higher-dimensional massive gravitons may be produced that could escape into the bulk, leading to events with 
missing energy in (3+1)-dimensional spacetime~\cite{Hamed1998,Mirabelli1999}.
Based on this premise the phenomenology of the SN 1987A supernova was investigated, obtaining limits on extra 
dimensions of $R < 3\times 10^{-6}$ m for $n=2$, $R < 4\times 10^{-7}$ m for $n=3$ and $R < 2\times 10^{-8}$ m for $n=4$~\cite{Cullen1999}.
Similarly, from a missing energy analysis of proton colliding events at the LHC, the constraint $R<3.2\times10^{-4}$ m can be extracted 
for $n=2$ imposing the $M_{4+n} = 1.93$ TeV bound~\cite{Aad2013}. Also for $n>2$, the bounds derived from the LHC are more stringent than 
those from molecules. However, additional assumptions beyond the ADD potential in Eq.~(\ref{VADD}) are necessary in order to interpret 
the LHC missing energy signals. 
Such assumptions are not needed for the molecular physics bounds, which are not sensitive to physics at very short distances.

The higher-dimensional Planck mass $M_{(4+n)}$ and corresponding Planck length $R_{\mathrm{Pl},(4+n)}$ derived from the present 
constraints on $R$ are also tabulated in Table~\ref{Constraints-R}. Note that in all cases the Planck energy scale 
in $(4+n)$ dimensions turns out to be in the range 1-500 MeV. Assuming a straightforward extrapolation of Eq.~(\ref{VADD}) 
from molecular distance scales to nuclear distance scales, for an effect to be observable at molecular physics scales, the gravitational 
interaction must grow so rapidly at smaller distances that gravity becomes strong and non-perturbative already in the MeV scale.
One would generally expect that strong gravity effects would already have been observed in scattering experiments, either through black hole 
creation, missing energy due to particles moving into the extra dimensions or other strong gravity effects, or large contributions 
to binding energies of nucleons and quarks. However, these expectations depend on a theory of gravity in the strong regime, 
and furthermore hadronic and nuclear binding is not a high-precision science, unlike atomic and molecular physics.
Here we take the attitude that we are constraining the possibility that the gravitational force has a higher-dimensional dependence 
on $r$ in the weak regime, where the computations are reliable, and we do not commit ourselves to what happens in the strong domain, 
where we cannot compute reliably. Our constraints are relevant only if gravity has a very soft behavior at shorter distances, 
so that its direct effects have escaped detection so far.

The idea that gravity at short distances may be soft has been explored in four dimensions in several contexts.
This includes asymptotically safe gravity \cite{Weinberg1979}, \emph{fat gravitons} as a possible solution to the
cosmological constant problem \cite{Sundrum2003}, a way of avoiding black holes and singularities in string 
theory~\cite{Siegel2003}, and bouncing or self-inflating universes \cite{Khoury2006}.
Therefore, given the limitations of our current understanding it is not unthinkable that this could happen in higher dimensions as well.

\section{Conclusion and outlook}

The alternative approaches for constraining compactification scales of extra dimensions, partially surveyed here, are all complementary as they 
probe different length and energy scales. Some approaches serve to produce tighter limits, however, often at the expense of additional assumptions.
In the present study, the constraints are derived from precision measurements on molecules, leading to straightforward interpretations.
Current state-of-the-art experiments on neutral hydrogen determine vibrational splittings of the order of $10^{-4}$ cm$^{-1}$, or 3 MHz~\cite{Dickenson2013}. 
Since the lifetimes of ro-vibrational quantum states in H$_2$ are of the order of $10^6$ s~\cite{Black1976}, 
measurements of vibrational splittings of the order of $10^{14}$ Hz could in principle be possible at more than 20-digit precision, 
which leaves room for improvement ``at the bottom" of over 10 orders of magnitude, if experimental techniques can be developed accordingly.

\ack

This presentation is based on work done in collaboration with Bert Schellekens, from the Institute of Subatomic Physics, Nikhef (Amsterdam), 
and Edcel Salumbides and Wim Ubachs, from the Vrije Universiteit of Amsterdam. I am very grateful to all of them.
The author is partially supported by funding from the Spanish Ministerio de Econom\' \i a y Competitividad, 
Research Project FIS2012-38816, and by the Project CONSOLIDER-INGENIO 2010, Programme CPAN (CSD2007-00042).

\end{document}